\documentclass[conference, 10pt]{IEEEtran}
\ifCLASSINFOpdf
\else
\fi
\usepackage{cite}
\usepackage{amsmath,graphicx,amssymb,color,textcomp,amsfonts,subfigure,enumerate,bm}
\usepackage{extarrows}
\usepackage{makecell,multirow,diagbox,stfloats}
\newcommand{\reffig}[1]{Fig. \ref{#1}}
\renewcommand\arraystretch{1.5}
\usepackage{amsthm}
\usepackage{algorithmicx,algorithm,balance}
\usepackage{algpseudocode}
\usepackage{colortbl}
\usepackage{array}
\usepackage{booktabs}
\usepackage{caption,siunitx}
\usepackage{forloop}
\newcounter{ct}


\begin{document}
\title{Zero-Power Backscatter Sensing and Communication Proof-of-Concept}

\author{
	Yu Zhang, Xiaoyu Shi, and Tongyang Xu\\
		School of Engineering, Newcastle University, Newcastle upon Tyne NE1 7RU, United Kingdom\\
	Email: \{yu.zhang, x.shi21, tongyang.xu\}@newcastle.ac.uk
}

\maketitle
\begin{abstract}
In this paper, we present an experimental setup to evaluate the performance of a radio frequency identification (RFID)-based integrated sensing and communication (ISAC) system. We focus on both the communication and sensing capabilities of the system. Our experiments evaluate the system's performance in various channel fading scenarios and with different substrate materials, including wood, plastic, wall, and glass. Additionally, we utilize radio tomographic imaging (RTI) to detect human motion by analyzing received signal strength indicator (RSSI) data. Our results demonstrate the impact of different materials and environments on RSSI and highlight the potential of RFID-based systems for effective sensing and communication in diverse applications.
\end{abstract}


\section{Introduction}
With the development of 6G communication technology, the role of integrated sensing and communication (ISAC) is expected to become increasingly significant. The integration of sensing and communication functions within a single device will create a more efficient and cost-effective hardware platform \cite{OpenJ}. One notable example of an ISAC system is the radio frequency identification (RFID) system, which has been widely used over the past decades. RFID systems utilize backscatter communication mechanisms to enable data exchange between a reader and tags without the need for a power source in the passive tags \cite{backscatter1}. Beyond its primary communication function, the RFID system also demonstrates sensing capabilities. The received signal strength (RSS) at the reader can vary depending on the environmental conditions and the positioning of the tags. By analyzing these variations in RSS, the system can realize different sensing functions, such as localization, tracking and motion detection.

Many studies have focused on the theoretical communication performance analysis of RFID systems. The work in \cite{backscatter2, modeling, bistatic} investigated the communication performance and modeling of RFID systems, exploring the fundamental limits and optimizing communication protocols to enhance efficiency and reliability. 
Beyond traditional modeling and optimization methods, work in \cite{transfer} proposed to use machine learning-based methods for backscatter communications, specifically employing deep transfer learning for signal detection in ambient backscatter systems. 
These studies provide valuable insights of how RFID systems can be designed and utilized for optimal communication performance.

On the experimental side, research has also been conducted to validate the sensing capabilities of RFID systems. \cite{dlf,multitarget} focused on using RFID experiments for sensing functions, such as motion detection and localization. These experiments demonstrate how RFID systems can detect and track motion and objects without requiring any devices to be worn or carried by the objects. This capability is particularly useful in scenarios where traditional sensing methods are impractical.

In this paper, we present an experimental setup to analyze both the communication and sensing functions of an RFID-based ISAC system. 
The experiment involves testing and evaluating the system’s communication performance in various channel fading scenarios and with different materials, such as tags attached to wood, plastic, walls, and glass. Additionally, we use a radio tomographic imaging (RTI) method to detect human motion based on the collected RSSI sensing data.
\section{Theoretical Model}
A typical RFID system consists of two primary components: a reader and a tag. In a passive RFID system, communication is realized through backscatter radio links. Unlike active tags, passive tags do not have an internal power source. Instead, they harvest energy from the continuous wave (CW) emitted by the reader through over-the-air links. The tag can modulate its own data onto the CW signal by altering the impedance of its antenna. This modulated signal is reflected back to the reader, known as backscattering. The reader subsequently demodulates this backscattered signal to retrieve the tag's ID, data, and received signal strength (RSS), which can be utilized for various sensing applications.

When multiple passive tags are present within the reader's field, the reader employs a collision avoidance mechanism called slotted ALOHA protocol, which is a part of the EPCglobal Class-1 Generation-2 (ISO 18000-6C) standard. This standard is widely used in the industry for ultra high frequency (UHF) RFID systems. The slotted ALOHA protocol works by dividing time into discrete slots and allowing tags to respond in randomly selected slots, thereby reducing the likelihood of signal collisions and ensuring that multiple tags can be read accurately and efficiently.
\subsection{Backscatter Communication Mechanism}
\textbf{Reader-to-Tag Communication}: In the proposed system, we utilize a commercial RFID reader (ThingMagic-M6e). The reader uses ASK modulation for the CW signals transmission and pulse interval encoding (PIE) for data encoding. ASK is a type of amplitude modulation that represents digital data as variations in the amplitude of a carrier wave. High amplitude corresponds to a transmitted CW signal, while low amplitude indicates an attenuated CW signal. PIE encodes data by varying the intervals between pulses. It uses a combination of short and long intervals to represent binary `0' and `1'.  For example, let $T$ represent a basic time unit, then a binary `0' might be encoded as a pulse of length $T$ followed by a space of length $2T$ and a binary `1' might be encoded as a pulse of length $2T$ followed by a space of length $T$. Since the reader provides energy to passive tags through these high-amplitude signals, PIE encoding is well-suited to the tags' power-harvesting capabilities \cite{rfidcommu}, \cite{rfidcommu2}.

\textbf{Tag-to-Reader Communication}: After receiving an ACK command from the reader, the tag responds with its electronic product code (EPC), a random number (RN16), a tag identifier (TID), and memory-stored data. The response is typically sent using backscatter modulation, where the tag modulates the reflection of the reader's CW signal to encode its data. The encoding scheme for the tag-to-reader link is determined by the reader, such as FM0 and Miller-modulated subcarriers (Miller-2, Miller-4, Miller-8).

\textbf{Channel Fading}:
The total backscattered power received by a monostatic reader is expressed by \cite{cascaded}
\begin{align}
y=\alpha \rho_LP_{TX}|G_TG_RL(d)\Gamma|^2|h|^4,
\end{align}
where $\alpha\in[0, 1]$ is the coefficient of tag's power transfer efficiency, $\rho_L$ is the polarization loss factor, $P_{TX}$ is the transmit power of reader, $G_T$ and $G_R$ are the antenna gains of tag and reader's antenna, respectively, $L(d)$ is the channel path loss model, $\Gamma$ is the differential reflection coefficient of the tag, and $h$ is the fading envelope of the channel link between the reader and tag. In the ideal line-of-sight (LOS) scenario, the $L(d)$ is given by
\begin{align}
L(d) =\frac{\lambda}{4\pi d},
\end{align}
where $\lambda$ is the communication signal wavelength, $d$ is the distance between reader and tag. However, in a real indoor environment, there might be multipath fading and interference, resulting in an RSS that is less than in the ideal LOS scenario.

\textbf{Substrate Materials}:
The RSS of passive tags can be affected not only by the channel fading environment but also significantly influenced by the substrate materials on which the tag is placed. In indoor environments, there are several typical materials to which tags can be attached, including metal, plastic, walls, and wood. The influence of these materials on RSS is different.

\subsection{RTI-based Sensing Mechanism}
\begin{figure}[ht]
	\centering
	\includegraphics[width=60mm]{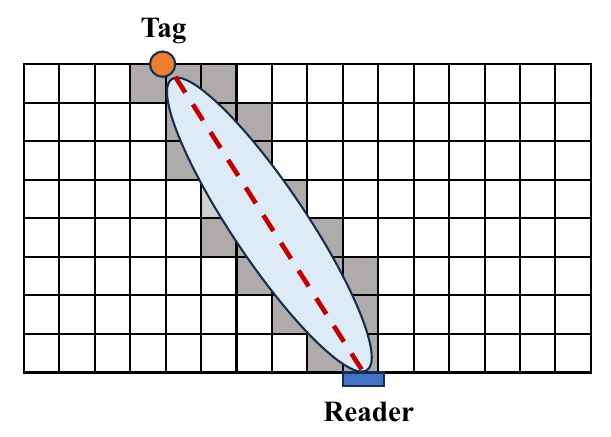}\\
	\caption{The link between a reader and a tag in an RTI network travels in a direct LOS path. The darkened grids represent the non-zero weight coefficients for this link.}\label{weightmodel}
\end{figure}
Radio tomographic imaging (RTI) leverages the attenuation differences in radio signals caused by obstacles, including human bodies. These differences are used to reconstruct images of the environment, allowing for the detection and localization of targets. Obstacles between the reader and the tag can absorb, reflect, and diffract radio signals, leading to changes in signal propagation. These changes provide valuable information about the environment and the location of objects. By analyzing the changed radio signals, a radio image of the environment can be reconstructed. This radio image represents the signal strength variations across different regions, which can be analyzed to monitor human motions and determine the location of a target.



By comparing the RSS measurements with the target present in the monitoring area to those taken when it is absent, we can identify the dynamic changes caused by the target. This process helps remove static losses. The shadowing loss caused by the target can be approximated as the sum of fading at all locations in the monitor area. The RSS change $\Delta y_i$ from time $t_1$ to $t_2$ can be expressed as \cite{RTI}
\begin{align}
\Delta y_i= \sum_{j=1}^{N} w_{ij}\Delta x_j + n_i,
\end{align}
where $w_{ij}$ is the weight coefficient of grid $j$ for link $i$, $x_j(t)$ is the signal attenuation value of grid $j$ at time $t$,  $\Delta x_j=x_j(t_2)-x_j(t_1)$, $N$ is the number of grids, and $n_i$ represents noise. The matrix form of all links can be expressed as
\begin{align}\label{eqY}
\textbf{y} = \textbf{W}\cdot \textbf{x} + \textbf{n},
\end{align}
where $\textbf{y}=[\Delta y_1, \Delta y_2, ..., \Delta y_Q]^T$ is the RSS variation of all $Q$ links, $\textbf{x}=[\Delta x_1, \Delta x_2, ..., \Delta x_N]^T$ is the RSS attenuations of all grids, $\textbf{n}=[ n_1, n_2, ..., n_Q]^T$ is the noise vector, and $\textbf{W}\in\mathbb{C}^{Q\times N}$ is the weight coefficient matrix.

We use an ellipsoid-based method to model the weight coefficients. As shown in \reffig{weightmodel}, if a grid falls outside the ellipsoid, its weight coefficient is set to zero. For grids within the path determined by the ellipsoid, the weight coefficient is inversely proportional to the square root of the link distance. For link $i$, the weight coefficient of grid $j$ is expressed as
\begin{align}\label{weight}
w_{ij}=
\begin{cases}
\frac{1}{\sqrt{d_i}}, &d_{ij}^R+d_{ij}^T<d_i+\beta,\\
0, &\text{otherwise},
\end{cases}
\end{align}
where $d_i$ is the length of the link $i$, $d_{ij}^R$ and $d_{ij}^T$ are the distances from the center of grid $j$ to the reader and the tag, respectively, and $\beta$ is a tunable parameter that is used for adjusting the width of the ellipsoid.

To obtain the RSS attenuations of all grids $\textbf{x}$ in \eqref{eqY}, we use the least square (LS) combined with Tikhonov regularization method \cite{RTI2}, which is solved by minimizing the noise power, as follows: 
\begin{align}
\hat{\textbf{x}}=(\textbf{W}^T\textbf{W}+\eta\textbf{C}^{-1})^{-1}\textbf{W}^T\textbf{y},
\end{align}
where $\eta$ is an adjustable regularization parameter, and $\textbf{C}$ is the covariance matrix, which can be expressed as
\begin{align}
\textbf{C}=\sigma^2\exp\left(-\frac{\|c_m-c_n\|}{\delta}\right),
\end{align}
where $\sigma^2$ is the variance of each grid value, $c_m$ and $c_n$ are the center coordinates of grid $m$ and grid $n$, respectively, $\|\cdot\|$ is the Euclidean distance, and $\delta$ is the correlation distance parameter.
\begin{table}[htbp]
\renewcommand{\arraystretch}{0.65}
\centering
\caption{RTI Parameter Configurations}
\begin{tabular}{ccc}
\toprule
Parameter&Description& Value\\
\midrule
 $\beta$& adjustable parameter & 10 \\
 $\eta$& regularization parameter & 1.5 \\
 $\sigma$& variance parameter&0.5 \\
$\delta$ & correlation parameter  & 3 \\
\bottomrule
\end{tabular}\label{tab:test}
\end{table}
\section{Experiment Design}

\begin{figure}[t]
	\centering
	\includegraphics[width=55mm]{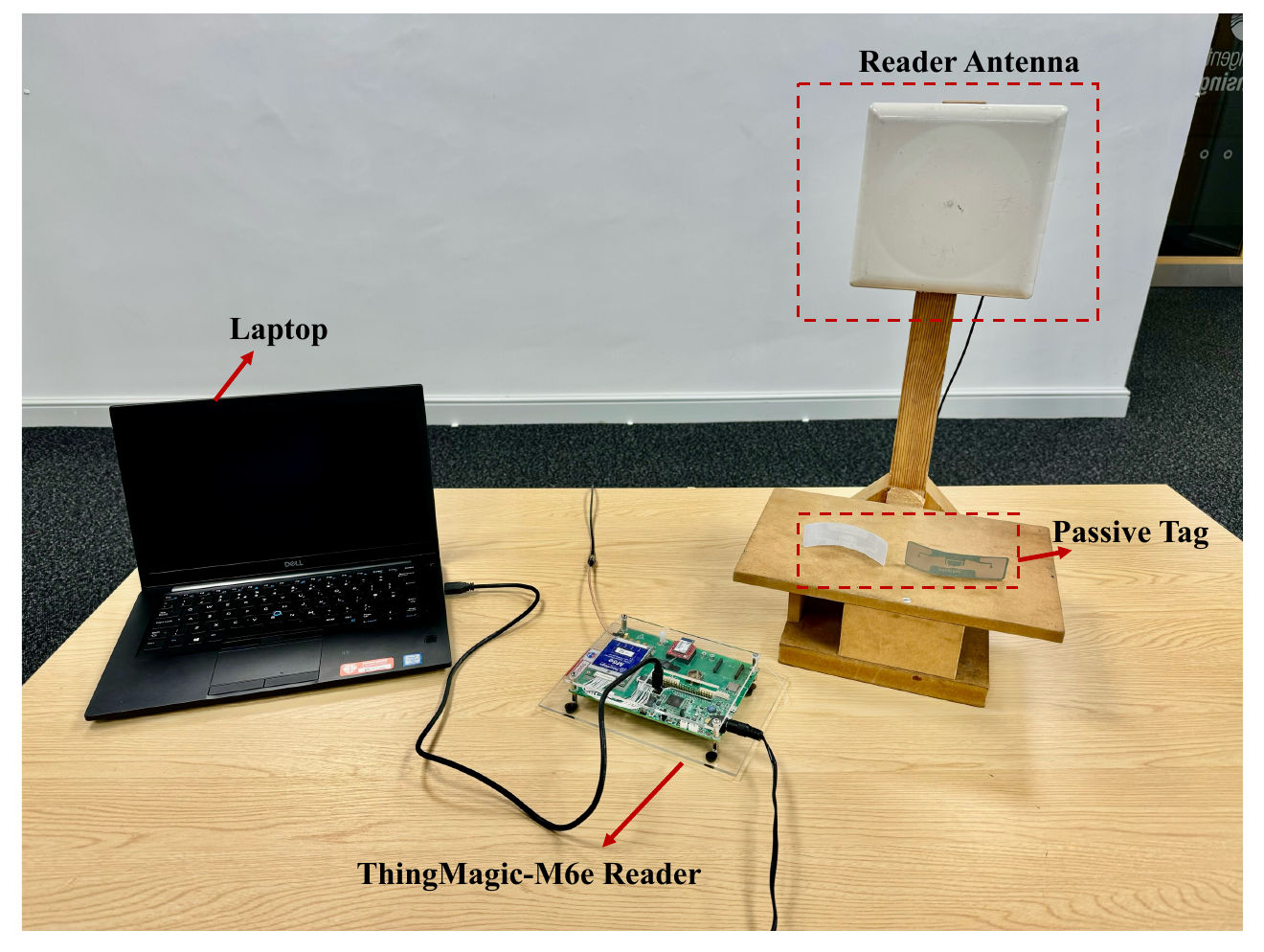}\\
	\caption{Experiment hardware: reader, antenna, and tag.}\label{hardware}
\end{figure}
\begin{figure}[h!]
	\centering 
	\subfigure[3D illustration.]
	{
		\includegraphics[width=0.6\linewidth]{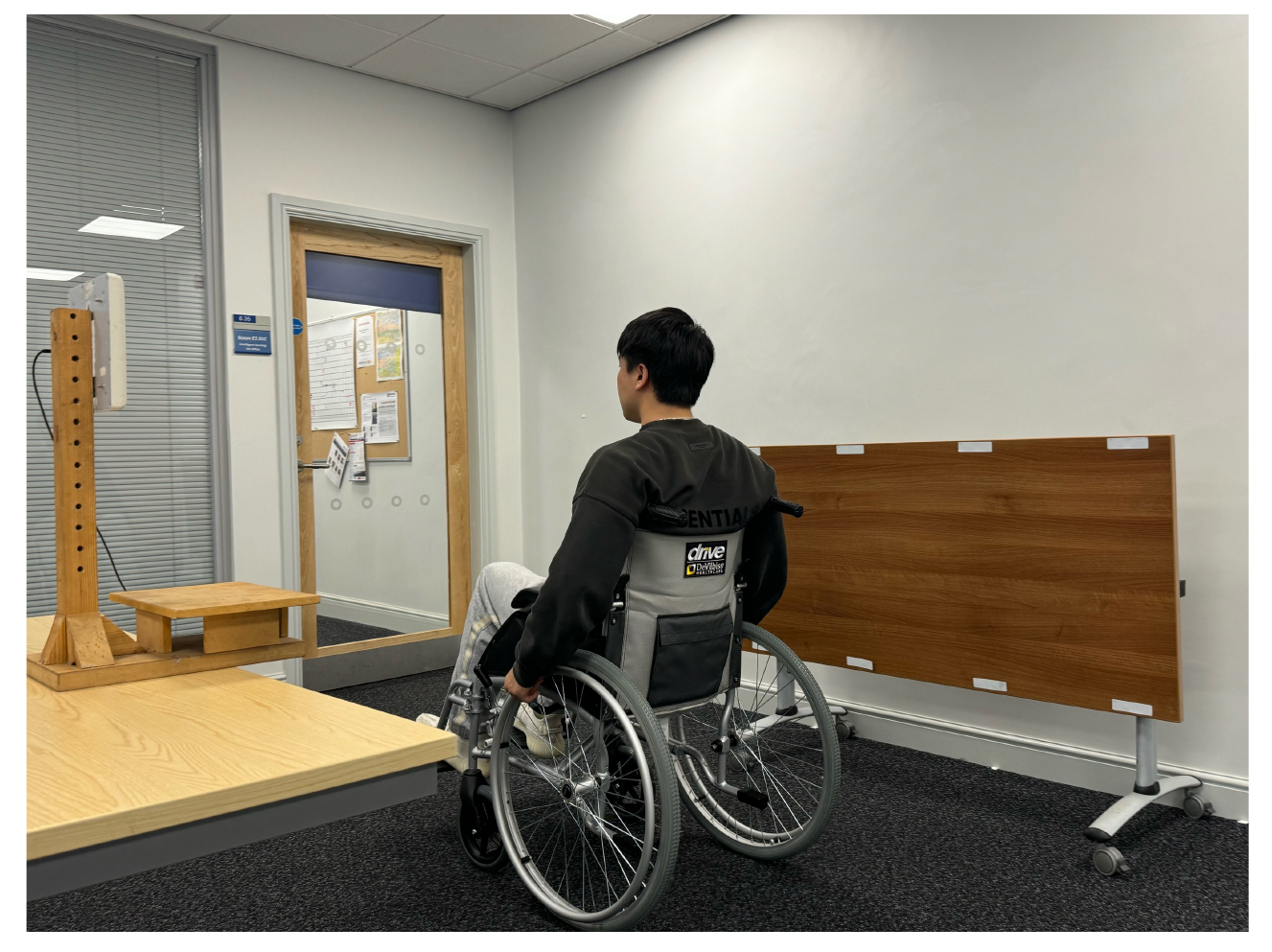}
		\label{3Dscenario}
	}
	\subfigure[2D illustration.]
	{
		\includegraphics[width=0.7\linewidth]{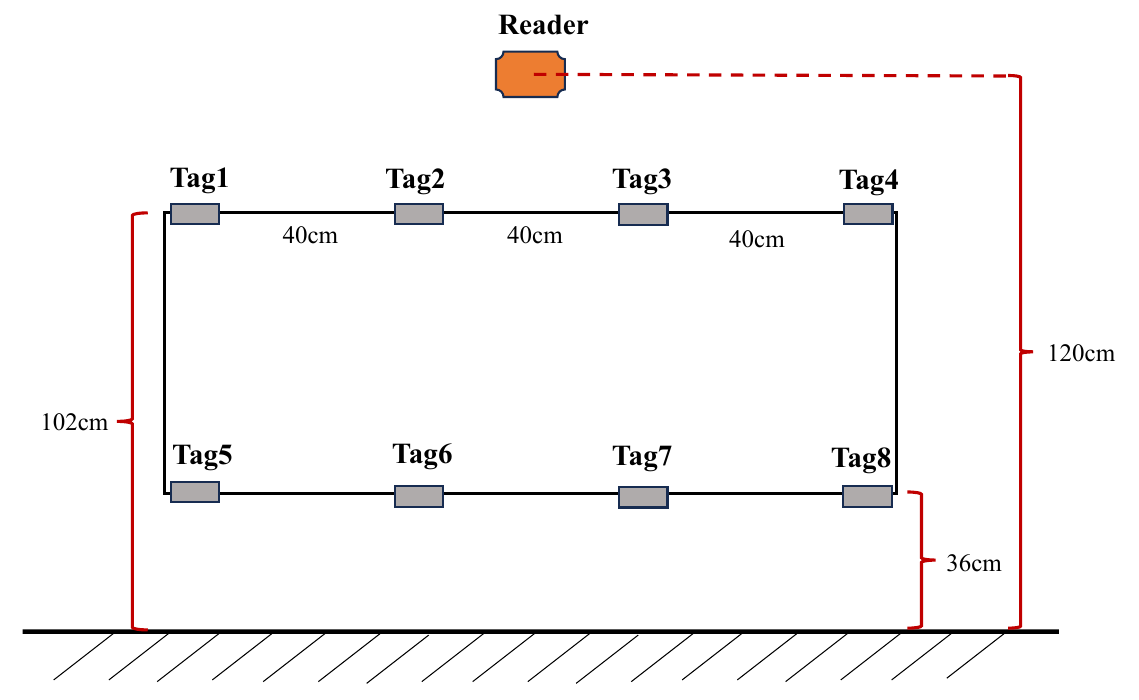}
		\label{2Dscenario}
	}
	\caption{Experiment setup for human motion detection.}\label{scenario}
\end{figure}

\begin{figure*}[h!]
	\centering 
	\subfigure[Poor-multipath LOS environment]
	{
		\includegraphics[width=0.25\linewidth]{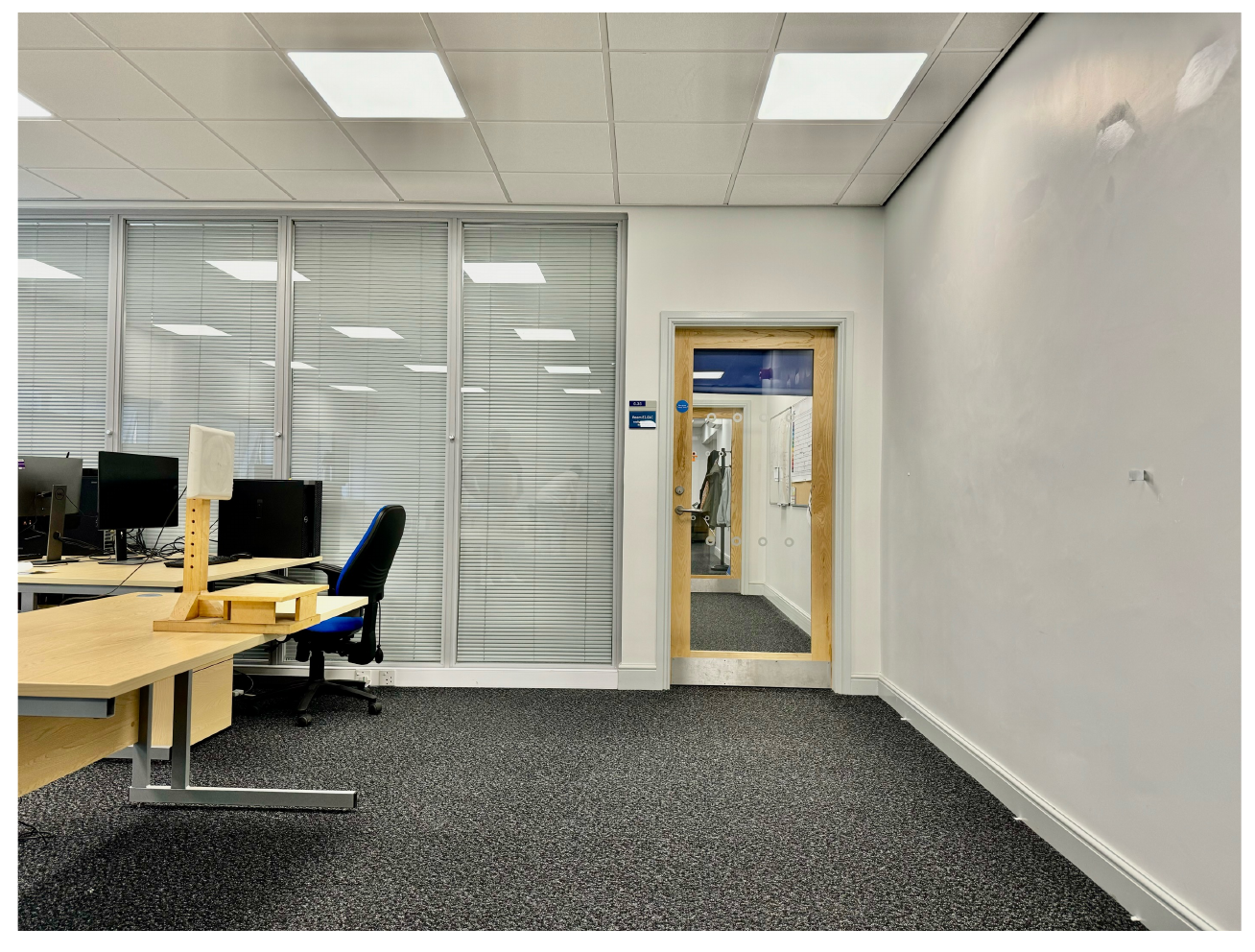}
		\label{LOS}
	}
	\subfigure[Poor-multipath NLOS environment]
	{
		\includegraphics[width=0.25\linewidth]{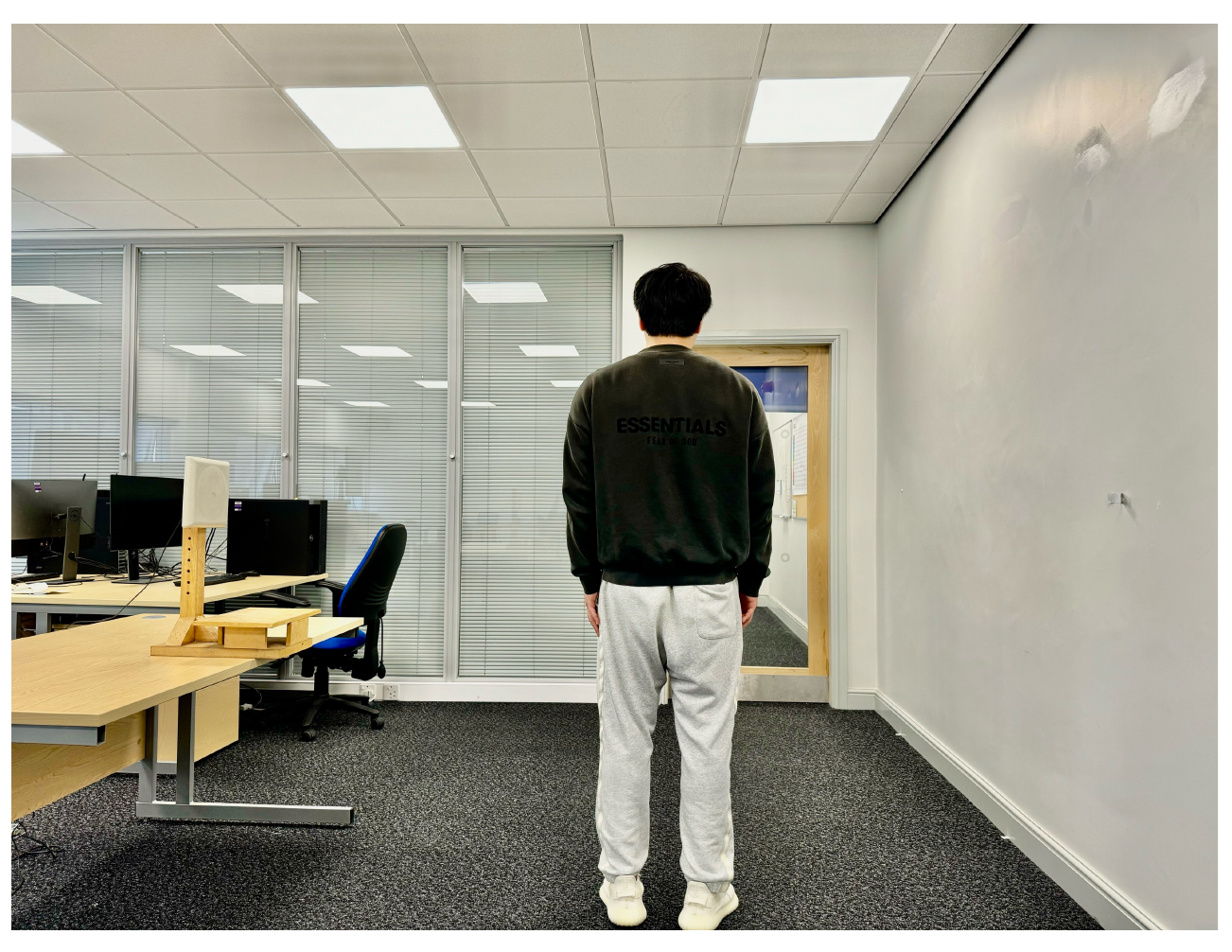}
		\label{NLOS}
	}
	\subfigure[Rich-multipath LOS environment]
	{
		\includegraphics[width=0.25\linewidth]{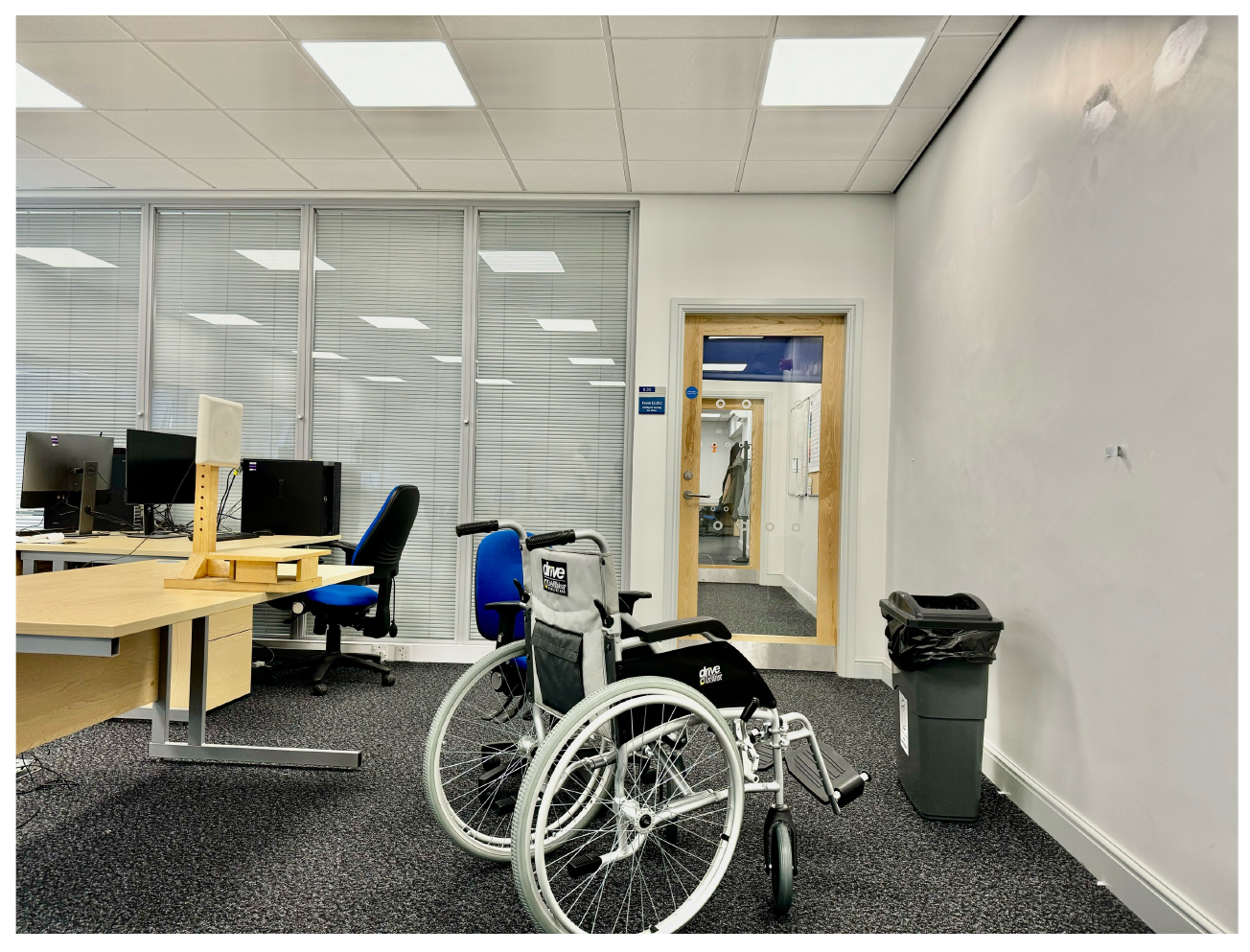}
		\label{multipath}
	}
	\caption{Experiment setup for different channel models. The tag could be attached to the wall or other materials. }\label{environment}
\end{figure*}
\begin{figure}[ht!]
	\centering
	\includegraphics[width=70mm]{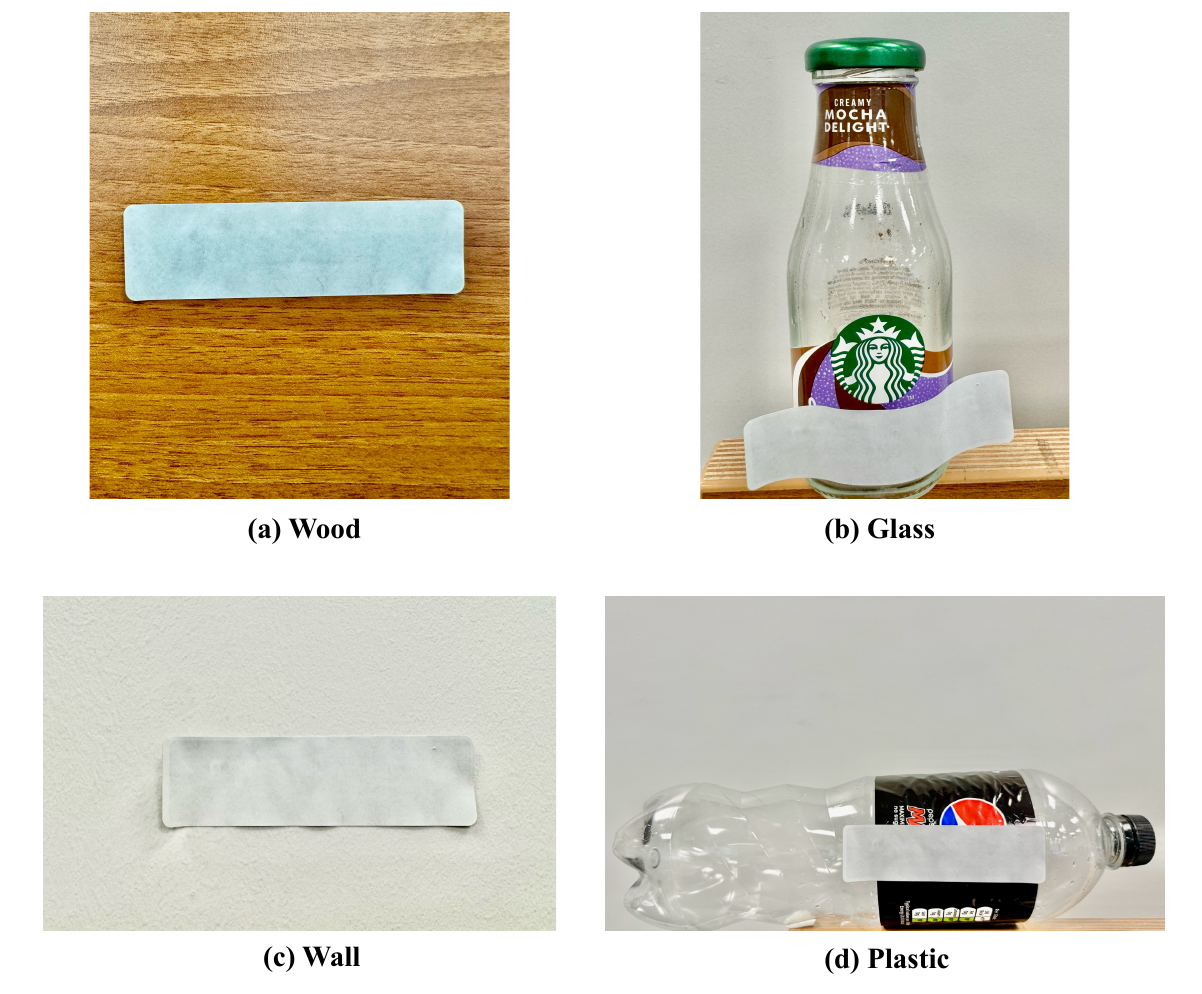}\\
	\caption{Tags attached to different materials.}\label{material}
\end{figure}
\subsection{Experiment Setup and RFID Devices}
In our experiment setup, we use a ThingMagic-M6e reader, an 865-956 MHz reader antenna with a 7.5 dB gain and a 3 dB elevation beamwidth of 72°, AD Dogbone R6 passive tags (size 97 x 27 mm), and a Windows laptop, as shown in \reffig{hardware}. The reader’s transmit power is set to 31.5 dBm, operating at a frequency of 902 MHz. We utilize the Universal Reader Assistant software to interface with the ThingMagic reader to obtain the tag’s RSSI, EPC, and data on the laptop. The RTI parameter configurations are presented in Table \ref{tab:test}.

As depicted in \reffig{scenario}, the horizontal distance from the center of the reader's antenna to the plane where the tags are located is 200 cm. The vertical distance from the center of the reader's antenna to the ground is 120 cm. The tags are spaced 40 cm apart. The distance from Tags 1-4 to the ground is 102 cm, while the distance from Tags 5-8 to the ground is 36 cm.

\subsection{Experiment Scenarios}
\subsubsection{Data Communications}
Passive data transmission via RFID is a fundamental function fully supported by the RFID hardware devices used in this work. Therefore, this experiment does not include this measurement.

\subsubsection{Human Motion Sensing}
As illustrated in \reffig{environment}, we evaluate RSS in three different channel fading scenarios \cite{RFIDAoA}, which can be used for human motion sensing. 
\begin{itemize}
	\item Poor-Multipath LOS Scenario: No obstacles block the direct path between the reader and the tags.
	\item Poor-Multipath NLOS Scenario: A human body blocks the path between the reader and the tags.
	\item Rich-Multipath LOS Scenario: Chairs, trash bins, and desks are randomly placed near the tags without blocking the LOS path.
\end{itemize}
\subsubsection{Substrate Materials}
As shown in \reffig{material}, we attach the tag to four different materials to test the influence of different substrate materials on RSS. The RSS variations can help optimize efficient communication links and can also be used to sense different materials. The substrate materials include wood (2.5 cm thick), a glass bottle (about 2 mm thick), a wall (about 11 cm thick), and a plastic bottle (0.25 mm thick).
\subsection{Data Collection}
{Scenario 1 - Baseline:} There are no movements or obstacles in the environment, and the duration is 60 seconds.

{Scenario 2 - Human Movement:}
One person continuously walks across the aisle between the reader and the tag at about 1 m/s, with the walking duration lasting about 10 seconds.


\section{Experiment Results}
\begin{figure}[ht!]
	\centering
	\includegraphics[width=70mm]{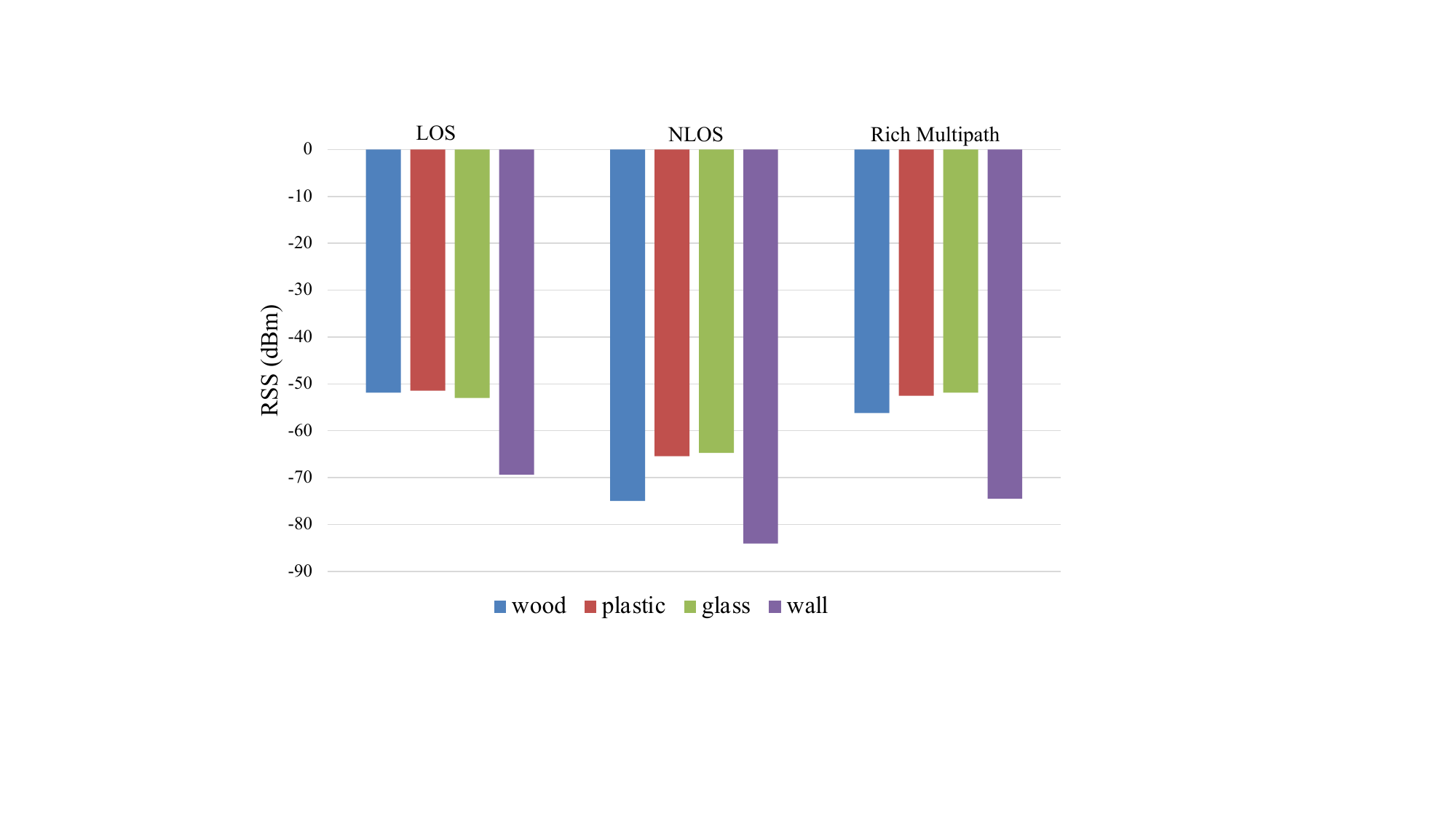}\\
	\caption{RSS values for tags attached to different materials.}\label{materialfig}
\end{figure}
\begin{figure}[ht!]
	\centering
	\includegraphics[width=90mm]{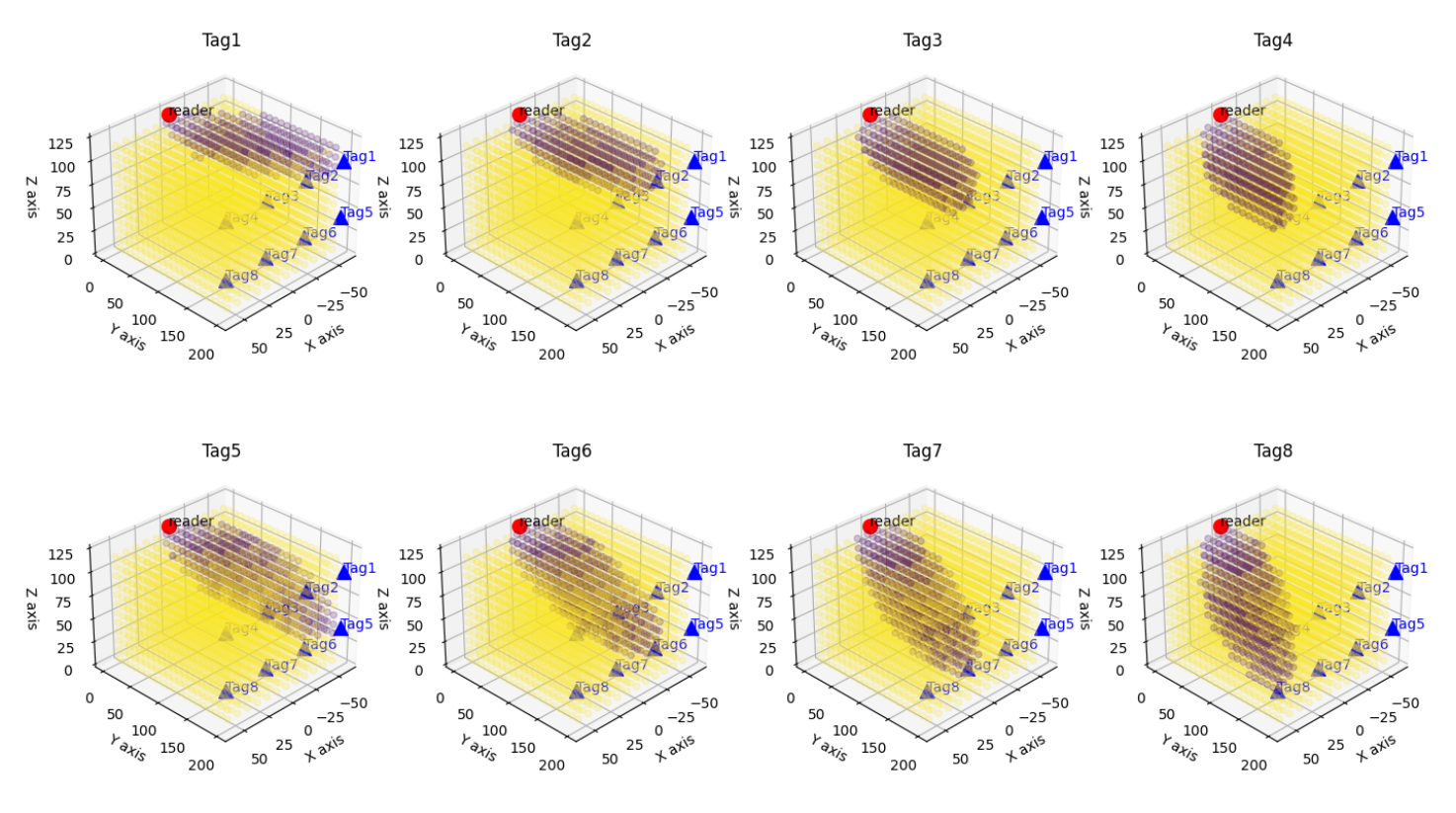}\\
	\caption{3D weight model for each reader-to-tag link.}\label{RSSlink}
\end{figure}

\begin{figure*}[h!]
	\centering 
	\subfigure[Baseline: no human motion at $t_0$]
	{
		\includegraphics[width=0.3\linewidth]{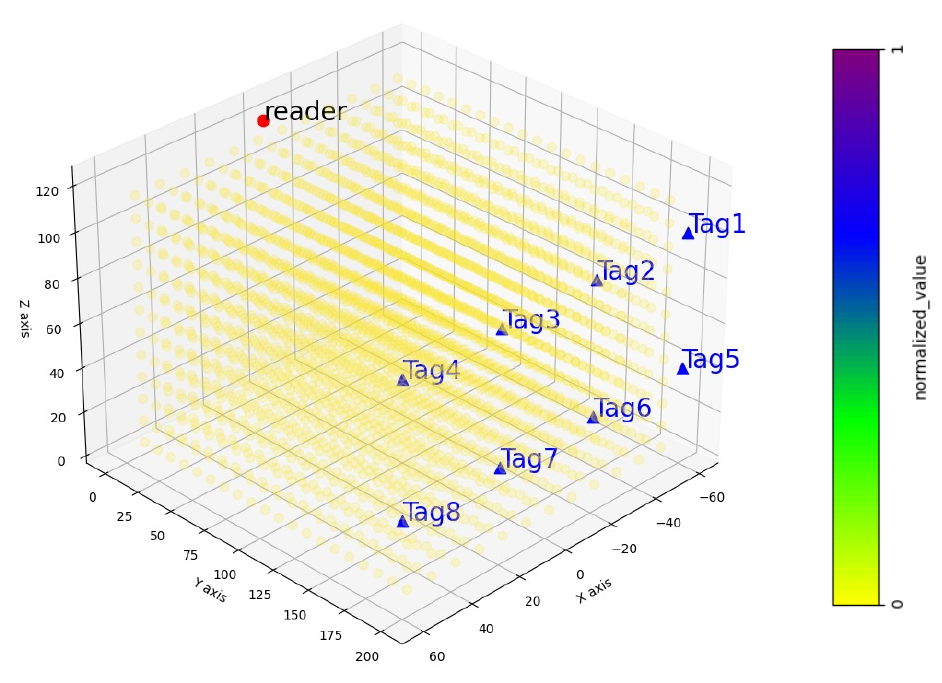}
		\label{motion0}
	}\\
	\subfigure[human motion at $t_1$]
	{
		\includegraphics[width=0.23\linewidth]{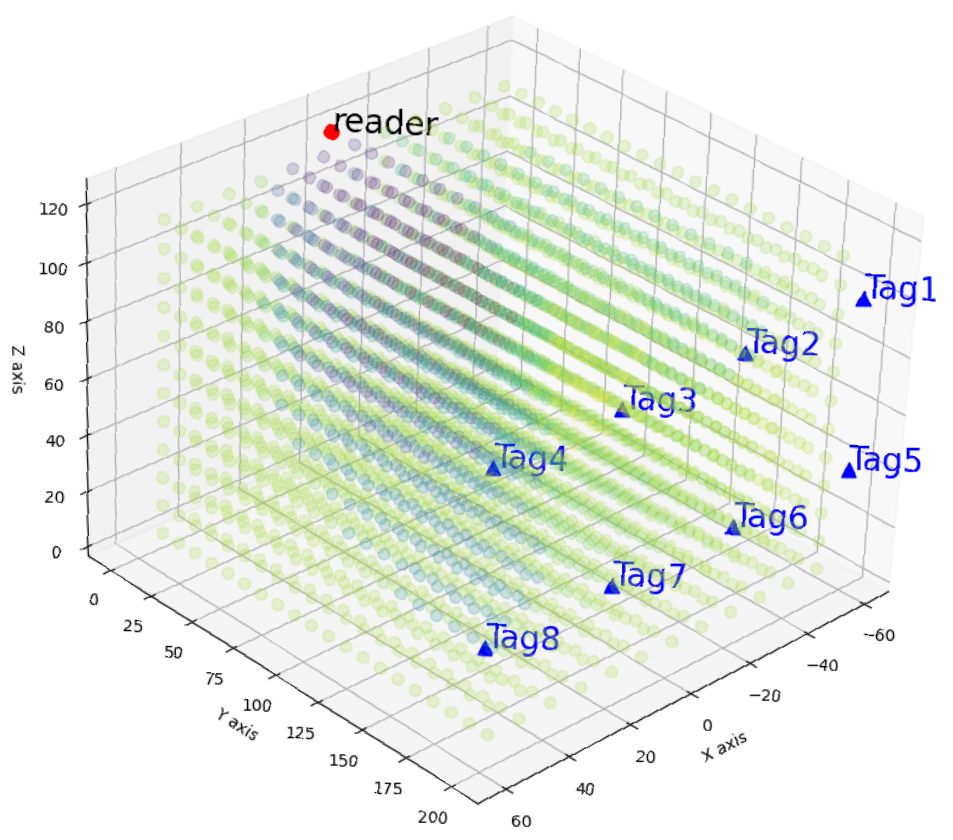}
		\label{motion1}
	}
	\subfigure[human motion at $t_2$]
	{
		\includegraphics[width=0.23\linewidth]{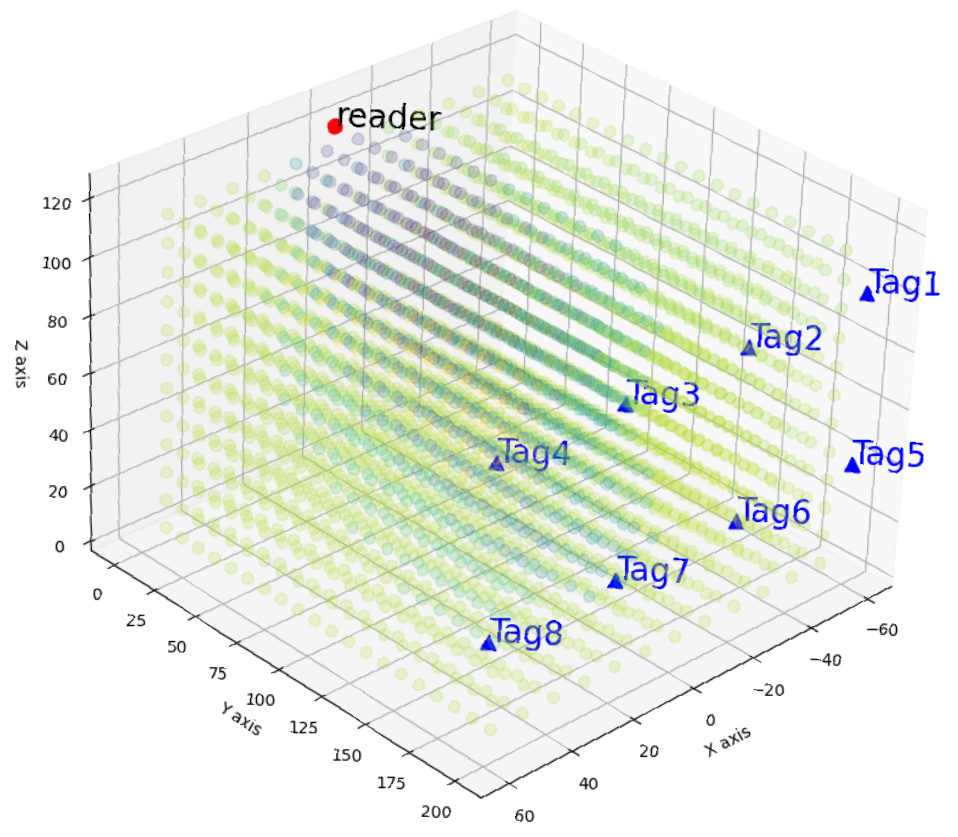}
		\label{motion2}
	}
	\subfigure[human motion at $t_4$]
{
\includegraphics[width=0.23\linewidth]{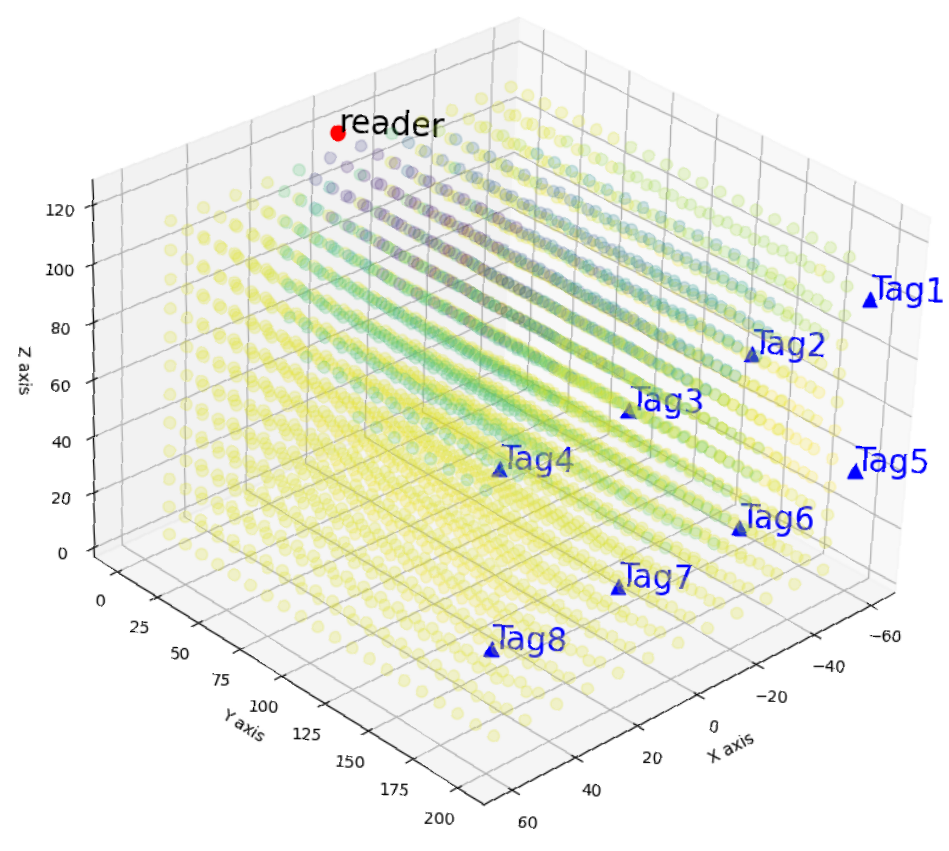}
\label{motion4}
}	
\subfigure[human motion at $t_5$]
{
\includegraphics[width=0.23\linewidth]{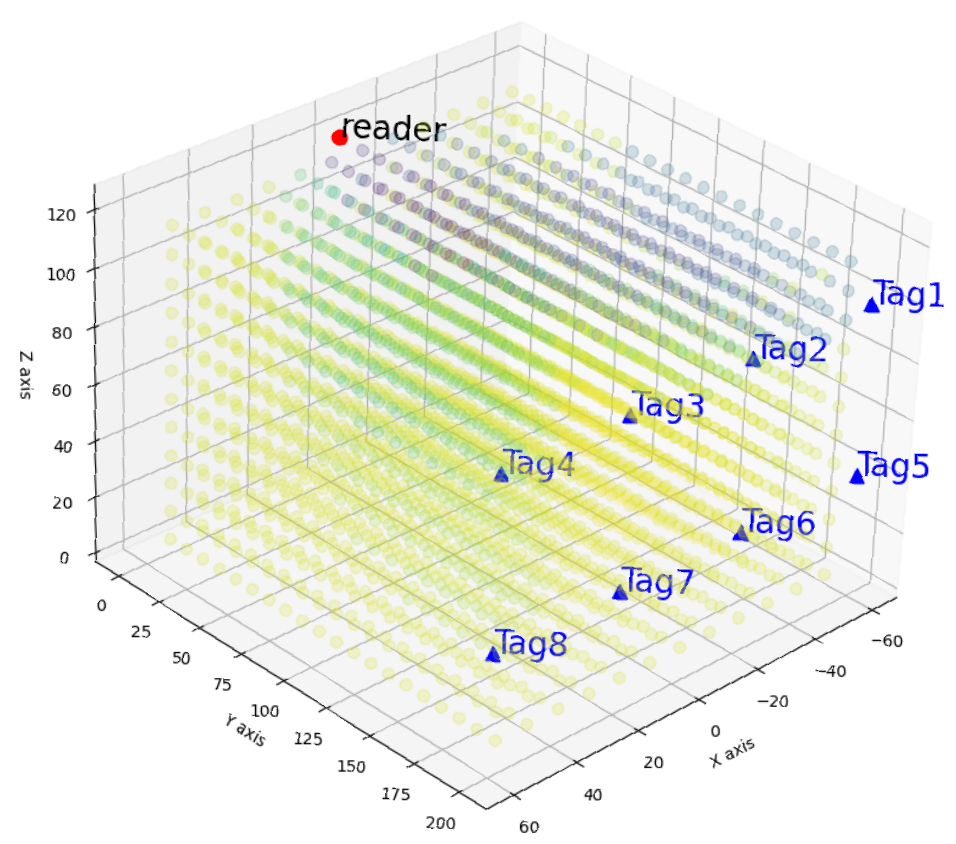}
\label{motion5}
}
\caption{Human motion detection: a person enters at tags 4,8 and leaves at tags 1,5 considering the \reffig{3Dscenario}'s practical scenario.}\label{motiondetection}
\end{figure*}

We conducted two main experiments. The first experiment focused on a static scenario where we attached tags to different materials, as shown in \reffig{material}. This experiment aimed to test the communication performance under different channel fading environments for various materials. The second experiment involved a human movement scenario, where we collected RSS data as a person moves across the tags to test the sensing functions. Here are the results of these experiments.

The results shown in \reffig{materialfig} were obtained by placing a single tag 200 cm horizontally away from the reader antenna. The tag was positioned 102 cm above the ground, and the reader antenna was 120 cm above the ground. The three different channel fading environments are illustrated in \reffig{environment}. As seen in \reffig{materialfig}, the RSS values are the worst when the tag is attached to the wall compared to when it was attached to wood, plastic, and glass. The RSS values for glass and plastic are almost identical. The best RSS values are observed in the LOS scenario, while the worst are in the NLOS scenario. Specifically, when the tag is attached to the wall, the RSS values drop to the tag's sensitivity threshold of -84 dBm, preventing the reader from reading the tag. These results indicate that for optimal communication performance, it is better not to attach the tag directly to the wall. The variations of RSS values in the NLOS scenario provide insights for sensing function analysis.

\reffig{RSSlink} presents the simulation results of the proposed weight model in \eqref{weight} for each reader-to-tag link. In the constructed 3D weight model, the darkened grids represent areas traversed by the reader-to-tag link, and yellow grids indicate areas not crossed by the link. \reffig{RSSlink} demonstrates the effectiveness of the RTI-based sensing method described in Section II.B.

\reffig{motiondetection} shows the RSS variations in the human motion scenario. In this experiment, a person enters near tag 4 and leaves near tag 1. Based on the RTI method, we can generate RSS variations at any time. Here we show the RSS variations at some representative time slots. \reffig{motion0} shows that at $t_0$, there is no human motion in the monitoring area, and the RSS shows no variations. \reffig{motion1} shows that at time $t_1$, a person walks by tag 4 and tag 8. \reffig{motion2} shows that at time $t_2$, a person walks by tag 3 and tag 7. \reffig{motion4} shows that at time $t_4$, a person walks by tag 2 and tag 6. Finally, \reffig{motion5} shows that at time $t_5$, a person walks by tag 1 and tag 5.
\section{conclusion}
Our experimental study demonstrates the dual capabilities of RFID-based systems for integrated sensing and communication, utilizing backscatter mechanisms. The findings reveal significant variations in RSS based on substrate materials and channel conditions, with notable performance degradation observed in NLOS scenarios, particularly when tags are attached to walls. The RTI method effectively detects human movement, showcasing the system's potential for motion sensing applications. These insights are crucial for optimizing RFID deployments in various environments, ensuring reliable communication and robust sensing capabilities. 

\section{Acknowledgement}
This work was supported by the UK Engineering and Physical Sciences Research Council (EPSRC) under Grant EP/Y000315/1.

 \end{document}